\documentclass[aps,prl,twocolumn,superscriptaddress,floatfix,longbibliography]{revtex4-2}
\usepackage[english]{babel}
\usepackage{amsmath,amssymb,bbm,mathrsfs,bm,braket,color,graphicx,comment,amsfonts,dsfont}
\usepackage{siunitx,physics}
\usepackage[colorlinks,citecolor=blue,urlcolor=blue]{hyperref}
\usepackage[mathscr]{euscript}
\usepackage{import}
\usepackage{cancel}
\usepackage{xcolor}
\usepackage{soul}
\begin{document}

\title{Interplay between inversion and translation symmetries in trigonal PtBi$_2$}

\author{Santiago Palumbo}
\affiliation{Instituto Balseiro, Univ. Nacional de Cuyo, Av. Bustillo, 9500, Argentina}
\author{Pablo S. Cornaglia}
\affiliation{Instituto Balseiro, Univ. Nacional de Cuyo, Av. Bustillo, 9500, Argentina}
\affiliation{Centro Atómico Bariloche, Instituto de Nanociencia y Nanotecnología (CNEA-CONICET), Av. Bustillo, 9500, Argentina}
\author{Jorge I. Facio}
\affiliation{Instituto Balseiro, Univ. Nacional de Cuyo, Av. Bustillo, 9500, Argentina}
\affiliation{Centro Atómico Bariloche, Instituto de Nanociencia y Nanotecnología (CNEA-CONICET), Av. Bustillo, 9500, Argentina}

\date{\today}

\begin{abstract}
The trigonal Weyl semimetal PtBi$_2$ presents an intriguing superconducting phase, previously reported to be confined to its topological Fermi arcs within a certain temperature range. This observation highlights the importance of a thorough understanding of its normal phase, particularly the roles that spin-orbit coupling (SOC) and inversion-symmetry breaking play in shaping its band structure. Our density-functional theory calculations reveal that the semimetallic nature of trigonal PtBi$_2$ can be interpreted as stemming from a noncentrosymmetric crystal distortion of a parent structure that drives a metal-to-semimetal transition. This distortion breaks inversion symmetry and, crucially, reduces translational symmetry. Due to its interplay with translational symmetry, inversion-symmetry breaking emerges as the dominant energy scale producing substantial asymmetries ($\sim$ 0.6\,eV) in certain short-range hopping amplitudes, superseding the effects of SOC, whose primary role is to define the characteristics of the low-energy nodal structure and of the topological Fermi arcs. This also applies to the formation of the Weyl nodes closest to the Fermi energy, which are found to exist even in the absence of SOC as a result of the orbital physics associated with the reduced translational symmetry.
\end{abstract}

\maketitle

\section{Introduction}

Broken inversion symmetry, combined with strong spin-orbit coupling, underlies a wide range of phenomena, from the Rashba effect to various nonlinear anomalous transport properties.
In its trigonal phase, PtBi$_2$ serves as a platform with these characteristics, showcasing a rich electronic structure. Notable features include three-fold band crossings~\cite{Gao2018,Jiang2020}, Rashba-like spin splitting~\cite{feng2019rashba}, and, more recently, novel signatures of its Weyl semimetallic nature~\cite{Veyrat2023,Kuibarov2023,hoffmann2024fermi,oleary2025topographyfermiarcstptbi2}.

The bulk and surface electronic structure of trigonal PtBi$_2$ have previously been studied by ARPES~\cite{Jiang2020,Kuibarov2023,PhysRevResearch.7.013025,oleary2025topographyfermiarcstptbi2}, quasiparticle interference~\cite{hoffmann2024fermi}, scanning tunneling microscopy~\cite{schimmel2024surface}, de Haas--van Alphen experiments~\cite{Gao2018,Veyrat2023}, and earlier band structure calculations~\cite{Gao2018,feng2019rashba,Kuibarov2023,PhysRevB.110.054504}.
This body of work has established key characteristics of the electronic structure such as the existence of various Fermi surfaces in the bulk of the system~\cite{Gao2018} as well as the relevance of the topological surface Fermi arcs to interpret different experimental observations~\cite{Kuibarov2023,hoffmann2024fermi,oleary2025topographyfermiarcstptbi2}.
The system has also attracted interest as a potential two-dimensional ferroelectric~\cite{PhysRevLett.133.186801} and the characteristics of the superconducting phase are a matter of ongoing experimental~\cite{Bashlakov2022,Veyrat2023,Kuibarov2023,schimmel2024surface,Zabala_2024,di2024interface} and theoretical research~\cite{PhysRevB.110.054504,Bai_2025,trama2024self}.
It is also a system of interest in the context of various anomalous transport responses~\cite{PhysRevB.110.125148,PhysRevB.109.235419,veyrat2024room,veyrat2024dissipationless}.

The intriguing properties of the superconducting phase naturally call for a detailed understanding of its normal phase.
A primary question of interest concerns the roles associated with $I$ symmetry breaking and with SOC. While the latter represents a large energy scale in Pt- and Bi-based compounds, e.g. relative to the superconducting gap, here we show that the $I$ symmetry breaking introduces an even larger scale in the system. This predominance originates in the nature of the crystal distortion, which in addition to breaking the $I$ symmetry, reduces the \textit{translational} symmetries present in the centrosymmetric limit.  This has profound effects in the electronic system. Viewed in a real space tight-binding picture, the reduction of translational symmetry allows the hoppings between near-neighbor orbitals to acquire sizable asymmetries, which we show can be significantly larger than the largest local SOC term. Viewed in momentum space, the larger unit cell of the $I$-broken system naturally leads to band folding effects. 

According to \textit{ab initio} calculations, PtBi$_2$ hosts two sets of Weyl nodes, located approximately 48\,meV and 150\,meV above the Fermi energy~\cite{Veyrat2023,PhysRevB.110.054504}. While the former are of primary interest, as their associated Fermi arcs have been observed through various experimental probes, understanding the origin of both sets provides valuable insight. We find that the Weyl nodes closest to the Fermi energy do not originate from nodal lines or Dirac-like crossings in the absence of SOC. Instead, the electronic structure in this limit features isolated band touchings that act as sources and sinks of the \textit{orbital} Berry curvature---that is, the Berry curvature arising from orbital and site degrees of freedom~\cite{lesne2023designing,mercaldo2023orbital}.

Due to SU(2) symmetry, these crossings carry a total Chern number of 2, one per spin channel. Upon inclusion of SOC, each such crossing splits into two nodes. One of them annihilates with a symmetry-related node, while the other survives in the real system.
In contrast, the Weyl nodes at 150\,meV emerge from mirror-protected nodal lines that become gapped by the introduction of SOC. Remarkably, for both sets of Weyl nodes, the nodal structure in the absence of SOC arises from band folding effects. These findings highlight the crucial role of the broken translational symmetries, which ultimately drive the orbital physics underlying the system’s nontrivial electronic topology.

Our discussions connect the electronic topology of PtBi$_2$ with paradigmatic phenomena where translational symmetry is spontaneously reduced, such as Peierls physics or long-range antiferromagnetic order, underscoring the roles that symmetry and topology can have to obstruct the development of a full gap for charge excitations in the distorted phase. The resulting picture also smoothly connects the physics of PtBi$_2$ with nontrivial electronic phases in honeycomb- and Kagome-like lattices. The general picture learned from these studies is provided in the accompanying paper Ref.~\cite{us_short} while  we focus here on its application in PtBi$_2$. 
This work is organized as follows. 
Section \ref{sec_methods} contains methodological aspects
 and presents the relevant symmetries of the problem.
Section \ref{sec_transition} describes the electronic reconstruction associated with the $I$-breaking distortion.
Section \ref{sec_wan} presents a microscopic analysis based on Wannier functions.
Section \ref{sec_nl} focuses on the origin of the different Weyl nodes in the system.
Section \ref{sec_fa} analyzes the impact of the spin-orbit coupling on the topological Fermi arcs.
Lastly, Section \ref{sec_conclusions} exhibits our concluding remarks while Appendix~\ref{sec_app} presents auxiliary phonon calculations.

\section{Methods}\label{sec_methods}

\subsection{\textit{Ab initio } calculations}

The electronic properties analyzed in this work are based on density-functional calculations (DFT) performed
with the FPLO package v22.01-63 ~\cite{Koepernik1999} based on the generalized gradient approximation (GGA)~\cite{Perdew1997} and the standard local basis setting as defined in \cite{lejaeghere2016reproducibility}. This setup has been previously found to describe well the experimental evidence on topological Fermi arcs in trigonal PtBi$_2$~\cite{Kuibarov2023,hoffmann2024fermi}. Input files are provided in \cite{palumbo_2025_15127964}. 

We construct Wannier Hamiltonians based on symmetry-preserved maximally projected Wannier functions as implemented in FPLO~\cite{koepernik23}. To this aim, we consider the Bi 6p and 6s orbitals as well as Pt 6s and Pt 5d orbitals. 
Integrations in the Brillouin zone (BZ) were performed using a tetrahedron method along with a $k$-mesh based on $19\times19\times17$ subdivisions of the first Brillouin zone.

\begin{figure*}[t]
    \centering
    \includegraphics[width=1.99\columnwidth]{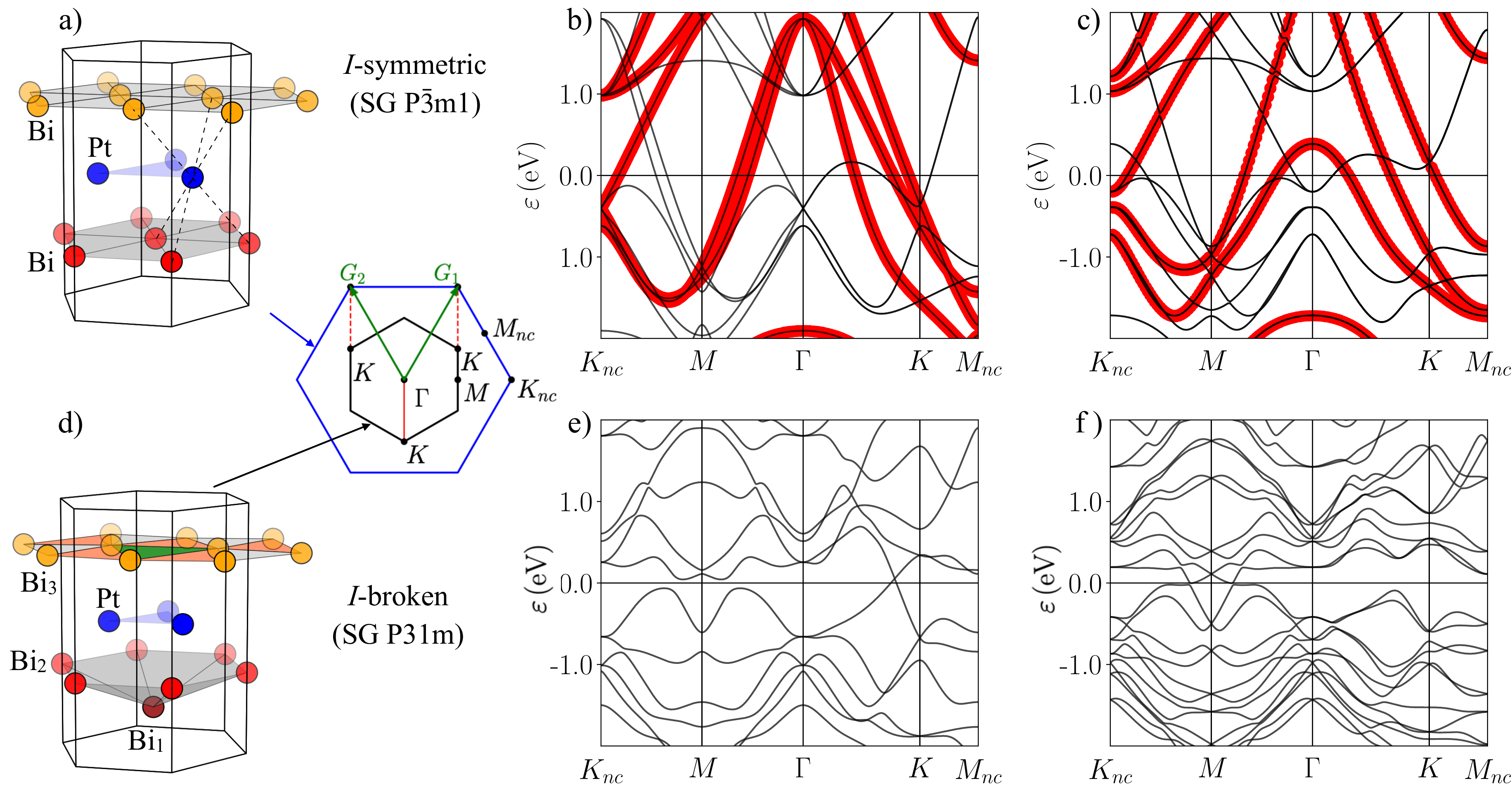}
    \caption{
(a) Crystal structure of trigonal PtBi$_2$ with space group $P\bar{3}m1$. This inversion-symmetric ($I$-symmetric) structure consists of stacked triangular lattices with Pt ions serving as inversion centers. The corresponding first Brillouin zone (FBZ) is outlined in blue, where the green arrows correspond to the primitive lattice vectors $G_i$. 
(b,c) Band structures calculated for the $I$-structure (b) without spin-orbit coupling (SOC) and (c) with SOC. Red dots indicate the nonfolded bands.
(d) Crystal structure of PtBi$_2$ in the inversion-broken structure ($P31m$). The corresponding FBZ is outlined in black. The red dashed segments outside the $I$-broken FBZ are translated by $G_1$ and $G_2$ to $\Gamma$-$K$. e,f) Band structures calculated for the $I$-broken structure (e) without SOC and (f) with SOC included.}
    \label{fig_dft}
\end{figure*}

\subsection{Crystal structure}

Fig. \ref{fig_dft}a,d) show, respectively, the crystal structure of PtBi$_2$ in the $I$-broken (space group P31m) and $I$-symmetric (space group P$\bar{3}$m1) phases~\cite{ptbi2:kaiser14,Shipunov2020}.
In the centrosymmetric case, the crystal consists of Bi and Pt layers, each forming a triangular Bravais lattice. 
The $I$-breaking distortion has a different effect on the two Bi layers. In one of them, the triangular lattice deforms into a decorated honeycomb lattice by pushing one out of every three Bi atoms outside the plane (Bi$_1$ in Fig. \ref{fig_dft}d). In the second Bi layer, Bi atoms are displaced along mirror-invariant planes so that sets of first neighbors no longer form identical equilateral triangles. Instead, chains of alternating large and small triangles can be observed along each of the preserved mirror symmetry planes.
Notice that the centrosymmetric structure has been experimentally obtained by moderate substitution of Bi by Te or Sb~\cite{Takaki2022}. 
In the repository~\cite{palumbo_2025_15127964}, we provide an interactive slider that allows the visualization of the structural distortion.
 
The main results presented in this work are obtained using the experimentally reported lattice parameters of noncentrosymmetric PtBi$_2$~\cite{Shipunov2020} and of centrosymmetric Pt(Bi$_{0.901}$Te$_{0.099}$)$_2$~\cite{Takaki2022}, presented in Table \ref{tab:structures}. Using fully relaxed internal coordinates leads only to modest changes in the energy and positions of the Weyl nodes~\cite{PhysRevB.110.054504}.

Throughout this work, we use the centrosymmetric structure as a reference, as it helps highlight the crucial role of the broken translational symmetries in noncentrosymmetric PtBi$_2$.
While we do not explore in detail the microscopic mechanisms by which Bi substitution with Te stabilizes the centrosymmetric phase observed in Ref.~\cite{Takaki2022}, in Appendix~\ref{sec_app} we discuss the dynamical stability of PtBi$_2$ in this phase.

In order to understand the consequences that the crystalline distortion has on the electronic structure, we perform DFT calculations for crystal structures that interpolate between the $I$-symmetric ($S_0$) and $I$-broken ($S_1$) cases. The interpolated structures are defined as 
 \begin{equation}
S_\alpha = (1-\alpha) S_0 + \alpha S_1,\,\,\, 0 \leq \alpha \leq 1,
\label{eq_interpolation}
\end{equation}
where the parameter $\alpha$ controls the degree of distortion. The interpolation involves both shifting the atomic positions and modifying the lattice parameters.
For this interpolation, the structure $S_0$ is represented in the same space group as $S_1$, which involves the construction of a $\sqrt{3}\times\sqrt{3}$ supercell.

In addition, for the $I$-broken structure $S_1$, we implement calculations varying the SOC. These are based on the linear interpolation
\begin{equation}
H_\lambda = (1-\lambda) H_0 + \lambda H_{SOC},\,\,\, 0 \leq \lambda \leq 1.
\label{eq_interpolation2}
\end{equation}
where $H_{SOC}$ and $H_0$ are the Wannier Hamiltonians obtained with and without SOC, respectively.

\begin{table}[t]
\centering
\caption{Crystal structures. $S_1$ is from Ref.~\cite{Shipunov2020} while $S_0$ corresponds to Pt(Bi$_{0.901}$Te$_{0.099}$)$_2$ in  Ref.~\cite{Takaki2022}. WP stands for Wyckoff position.}
\label{tab:structures}
\begin{tabular}{|l|l|l|l|l|}
\hline
$S_1$         & SG P31m (157)                          & $a$=6.57316\AA & $c$=6.16189\AA &         \\ \hline
WP & Atom                                   & $x/a$         & $y/a$         & $z/c$       \\ \hline
3c..m            & Pt                                     & 0.2619    & 0         & 0.004   \\ \hline
1a3.m            & Bi$_1$                                     & 0         & 0         & -0.359 \\ \hline
2b3..            & Bi$_2$                                     & -1/3      & 1/3      & -0.204 \\ \hline
3c..m            & Bi$_3$                                     & -0.3856   & 0         & 0.271  \\ \hline
\hline 
$S_0$         & SG P$\bar{3}$m1 (164) & $a$=4.06388\AA & $c$=5.52653\AA &         \\ \hline
WP & Atom                                   & $x/a$         & $y/a$         & $z/c$       \\ \hline
1$\bar{3}$m.     & Pt                                     & 0         & 0         & 0       \\ \hline
2d3m.            & Bi                                     & 1/3       & -1/3      & 0.25468  \\ \hline
\end{tabular}
\end{table}

\subsection{Band unfolding}

Since the $I$-broken unit cell corresponds to a supercell of the $I$-symmetric system, the first Brillouin zone of the former is correspondingly smaller (Fig.~\ref{fig_dft}). In this context, it is useful to characterize how the bands in one setup are folded or unfolded into the other.  
To this end, we employ the unfolding methodology implemented in FPLO~\cite{PhysRevLett.106.027002}.  
This method, building on ideas by Wei Ku {\it et al.}~\cite{unfolding}, introduces the so-called unfolding weight, which enables an exact distinction between folded and nonfolded bands in the limit where a supercell is constructed but no perturbation reducing translational symmetry is introduced. 
In this limit, the unfolding weight vanishes for folded Bloch states and equals one for nonfolded states.
For a distorted structure, this quantity adopts intermediate values which provide a means to track the mixing between folded and nonfolded bands induced by the distortion.  For a complete description, we refer to the FPLO documentation~\cite{Koepernik1999}.

\section{Metal to semimetal transition}
\label{sec_transition}

We begin our description of the electronic structure in the scalar-relativistic approximation in which the SOC is neglected. We shall see in this paper that the SOC is important to delineate the characteristics of the band touchings near the Fermi energy but that the $I$-breaking distortion operates in a larger energy scale, controlling a metal to semimetal transition.

Fig. \ref{fig_dft}b) presents the band structure obtained for the $I$-symmetric case in the absence of SOC. To ease the comparison with the $I$-broken case, shown in Fig. \ref{fig_dft}e), we have performed both calculations representing the system with the noncentrosymmetric space group. In addition to the bands which are depicted with lines, the plot also includes dots whose size is proportional to the unfolding weight, which exactly distinguishes folded from nonfolded bands in the $I$-symmetric limit. 

The particularly notable fourfold crossing (excluding spin) at the $\Gamma$ point with an energy of approximately $-0.4$\,eV corresponds to states originally at the $K$ points of the centrosymmetric BZ (indicated as $K_{nc}$ in Fig.~\ref{fig_dft}). The set of bands involved in this crossing is of interest for two reasons. Firstly, because they vividly illustrate the strong impact of the $I$-breaking distortion in this material: upon $I$ symmetry breaking, these bands split into two crossings separated by $\sim0.8$\,eV, one above and one below the Fermi energy, as can be seen in Fig. \ref{fig_dft}e).
Secondly, as we show in Section~\ref{sec_nl}, the nontrivial electronic topology of PtBi$_2$ essentially arises from topological touchings induced by the crystalline distortion between the occupied set of these folded split bands with nonfolded bands. 

Figs.~\ref{fig_dft}c,f) show the band structure for the $I$-symmetric and $I$-broken systems, respectively, including the SOC. The distinction between a metallic phase in the former and a semimetallic phase in the latter remains evident in this fully relativistic case as well.
In addition to inducing a substantial Rashba splitting at the $M$ point--consistent with Ref.~\cite{feng2019rashba}--the main effect of the SOC near the Fermi energy is to modify the nodal structure. In particular, the band crossing near the Fermi level along $\Gamma$–$K$ becomes gapped. These effects are analyzed in detail in Section~\ref{sec_nl}.

\begin{figure}[t]
    \centering
    \includegraphics[width=\columnwidth]{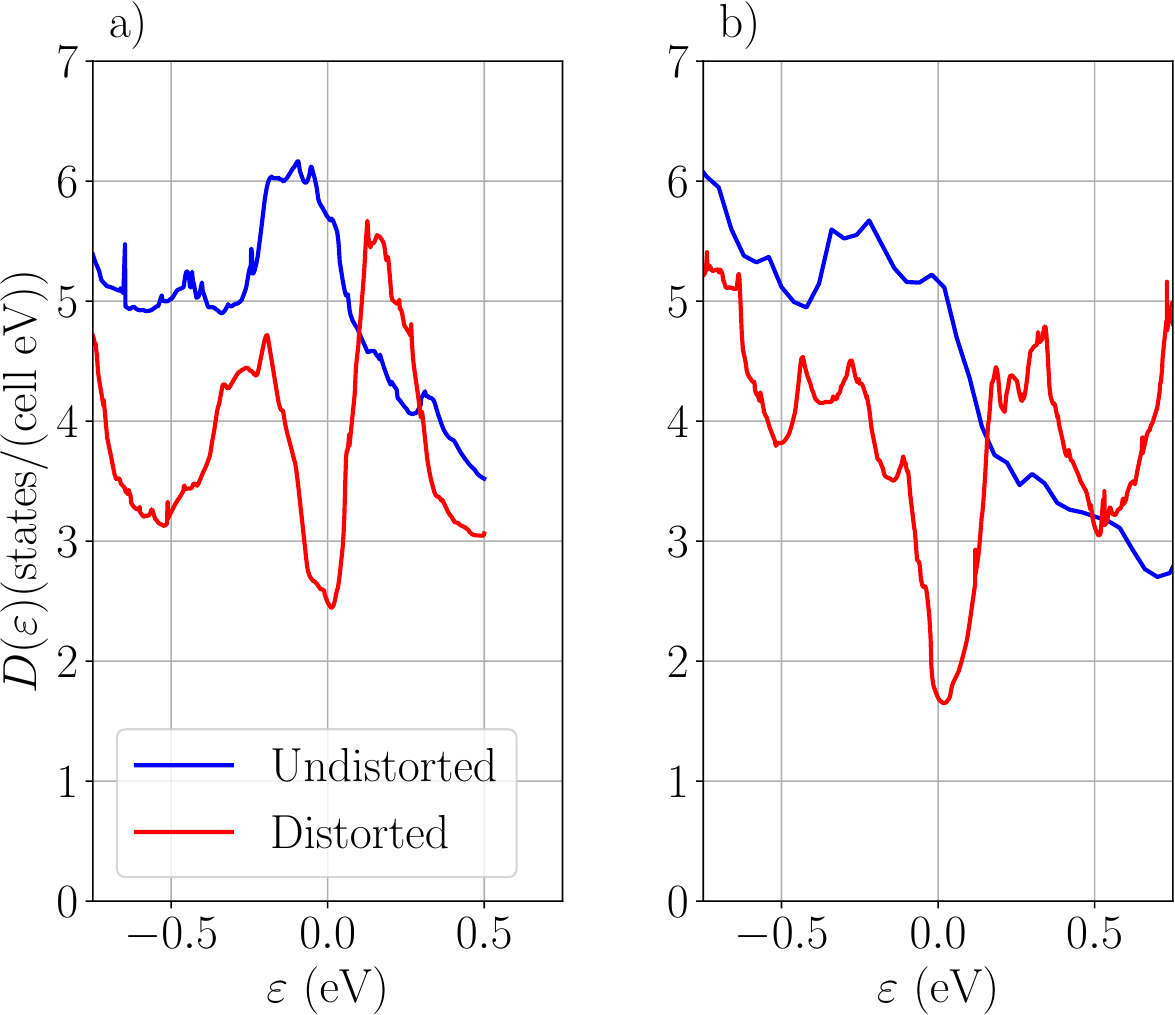}
    \caption{Density of states without (a) and with SOC (b). }
    \label{fig_dos}
\end{figure}

The opening of a substantial gap at $\Gamma$ signals a strong electronic reconstruction that removes significant spectral weight near the Fermi energy. This effect is evident in the density of states (DOS), which decreases at the Fermi level from approximately 6 states/(eV cell) to nearly 2 states/(eV cell) as $I$ symmetry is broken (Fig.~\ref{fig_dos}). In the $I$-broken case, the DOS near the Fermi energy develops a pseudo-gap, characteristic of semimetals, supporting the interpretation of a metal-to-semimetal transition that we adopt throughout this work. 

\begin{figure*}[t]
    \centering
    \includegraphics[width=2\columnwidth]{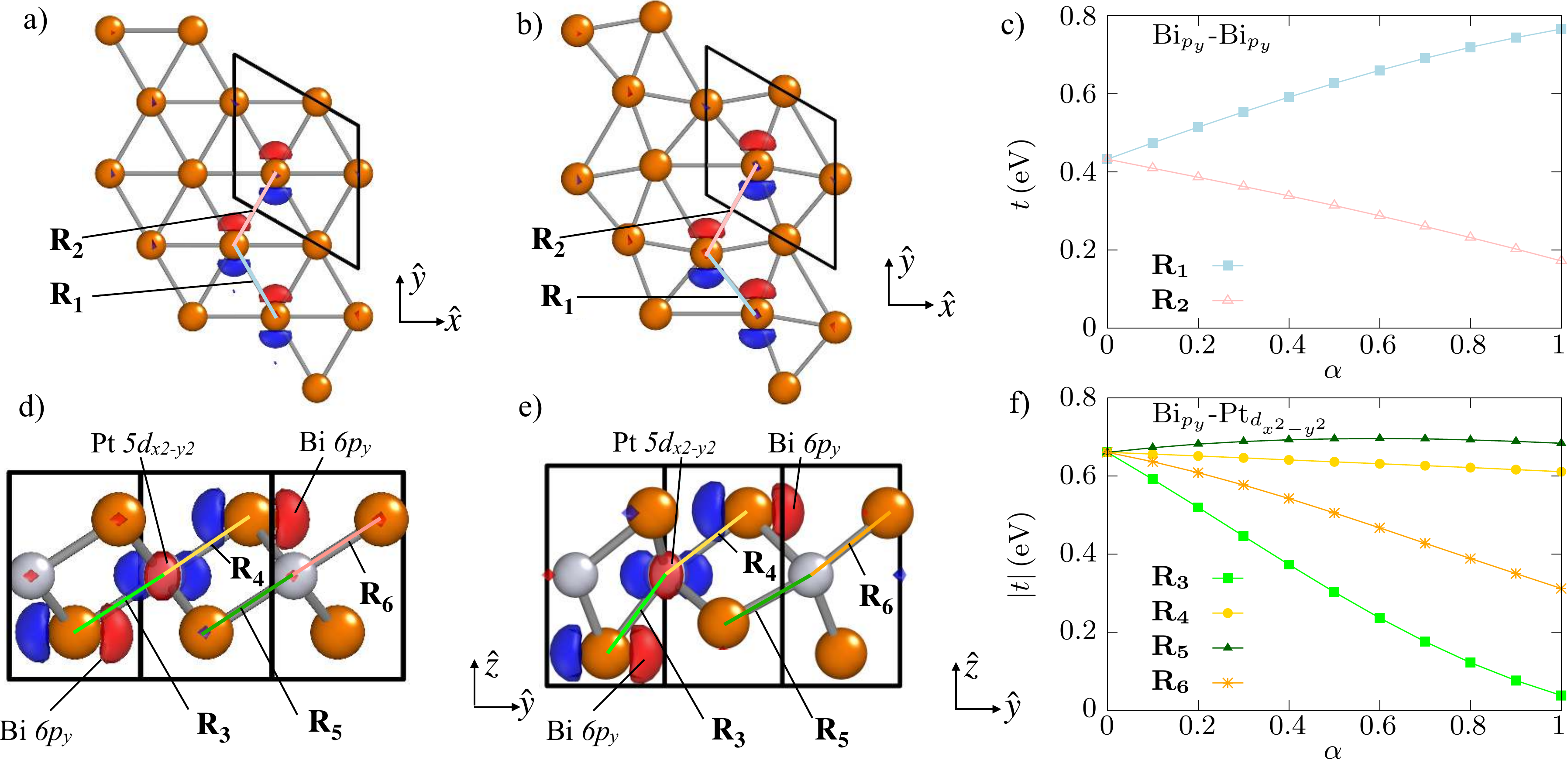}
    \caption{a,b) Top view of the upper Bi plane for the $I$-symmetric and $I$-broken structures, respectively. The vectors $\mathbf{R_1}$ and $\mathbf{R_2}$ have distinct lengths in the noncentrosymmetric case. c) Amplitude of the matrix elements between Bi 6$p_y$ orbitals as a function of the crystalline distortion [$\alpha$ in Eq. (\ref{eq_interpolation})] along bonds $\mathbf{R_1}$ and $\mathbf{R_2}$ illustrated in panels a,b. d,e) Side view of the undistorted and distorted crystal structures, highlighting four bonds which become inequivalent in the latter case. f) Amplitude of the hopping along these four bonds between Bi 6$p_y$ and Pt 5$d_{x^2-y^2}$ orbitals as a function of the crystalline distortion parameter $\alpha$.}
    \label{fig_wan}
\end{figure*}

\section{Wannier-functions-based analysis}
\label{sec_wan}
We have demonstrated that the crystalline distortion breaking inversion symmetry introduces substantial modifications to the band structure, including a splitting of 0.8\,eV at $\Gamma$ (Fig. \ref{fig_dft}). To better characterize the microscopic origin of these effects, we construct Wannier functions and analyze the evolution of the corresponding matrix elements as the distortion is introduced. To this aim, we consider the interpolation between crystal structures defined in Eq. \ref{eq_interpolation}.
As a reference, we also compute Wannier functions for the fully relativistic case to establish the energy scale of the SOC. Due to the $C_{3v}$ point symmetry, two independent parameters describe the local SOC for Bi $6p$ orbitals: $|\langle p_{x\uparrow} | H | p_{y\uparrow} \rangle| = 0.36$\,eV and $|\langle p_{x\uparrow} | H | p_{z\downarrow} \rangle| = 0.3$\,eV~\cite{varjas2018qsymm}.

Among the non-relativistic matrix elements that are equal in amplitude in the $I$-symmetric limit, we identify several instances where the asymmetry induced by $I$ symmetry breaking is comparable to or even exceeds the local SOC terms. Figure~\ref{fig_wan} highlights two cases where this effect is particularly pronounced and that help to clarify from a microscopic point of view the interplay between broken inversion and translational symmetries.

Fig. \ref{fig_wan}a,b) focus on the hopping between Bi $6p_y$ Wannier functions within the upper Bi plane. The distortion splits the first-neighbor bonds of the triangular lattice into short and long bonds. In the distorted structure, the hopping along the shorter bond (indicated by $\mathbf{R_1}$ in Fig. \ref{fig_wan}b) increases by 0.6\,eV relative to the longer bond ($\mathbf{R_2}$), as shown in Fig. \ref{fig_wan}c).
This asymmetry reflects the reduction of the translational symmetry: in the $I$-symmetric limit, these two bonds are connected by the combined action of a reflection ($x \to -x$) and a translation by a lattice vector of the undistorted triangular lattice.

We note that in simple tight-binding models with $s$--like orbitals at the Bi atomic sites, the introduction of this kind of asymmetry in the first-neighbor hoppings leads to a metal to semimetal transition~\cite{us_short}.

The second case, illustrated in Fig.~\ref{fig_wan}d,e), involves the hopping between Pt $5d_{x^2-y^2}$ and Bi $6p_y$ orbitals from the Bi layers above and below the Pt layer.  
In the $I$-symmetric structure, these hopping amplitudes are equal because the Pt atoms act as inversion centers.  
In the $I$-broken structure, where each unit cell contains three Pt sites, three pairs of bonds can exhibit asymmetries.  
Figure~\ref{fig_wan}d) shows two such pairs: $\{\mathbf{R_3}, \mathbf{R_4}\}$ and $\{\mathbf{R_5}, \mathbf{R_6}\}$. The third pair is not shown, as it is related to $\{\mathbf{R_5}, \mathbf{R_6}\}$ by reflection symmetry. Asymmetries of approximately 0.5\,eV are observed [see Fig.~\ref{fig_wan}f)]. Notably, some of these asymmetries, such as the difference in hopping amplitudes along $\mathbf{R_3}$ and $\mathbf{R_5}$, are also related to the reduction in translational symmetry, as these bonds are connected by a lattice translation in the undistorted case.
 
\section{Weyl nodes and their origin}
\label{sec_nl}

The existence of Weyl nodes is often understood as the result of perturbations acting on preexisting fourfold crossings. A well-known example is the splitting of isolated fourfold crossings by an external magnetic field~\cite{PhysRevLett.111.246603,PhysRevB.88.165105,hirschberger2016chiral,PhysRevB.98.075123}.
A second example arises from a theoretical framework that considers the effects of SOC separately. In several compounds, such as NbP~\cite{PhysRevX.5.011029} and Co$_3$Sn$_2$S$_2$~\cite{liu2018giant}, theoretical studies have shown that, in the absence of SOC, the electronic structure features nodal lines, namely, continuous curves where four bands (counting spin) intersect. These nodal lines are typically found in reflection-invariant planes~\cite{PhysRevX.5.011029,liu2018giant,10.1063/1.5123222}, with the crossing bands carrying opposite mirror symmetry. SU(2) symmetry protects these nodal lines by preventing coupling between states with the same mirror symmetry but opposite spin. When SOC is introduced, breaking SU(2) symmetry, the nodal lines gap out except at isolated band touchings, which become Weyl nodes. In other systems, such as WTe$_2$~\cite{soluyanov2015type} and  TaIrTe$_4$~\cite{PhysRevB.93.201101}, the electronic structure hosts Weyl nodes even in the absence of SOC. In that limit, the Berry curvature giving rise to the finite Chern number arises from orbital and site degrees of freedom~\cite{lesne2023designing,mercaldo2023orbital}.

In PtBi$_2$, we focus on the Weyl nodes that connect bands $N$ and $N+1$, with $N$ the number of valence electrons. 
Within an energy window of 200\,meV around the Fermi energy, PtBi$_2$ hosts two sets of Weyl nodes, located at approximately 48\,meV (set I) and 150\,meV 
(set II) above the Fermi energy~\cite{Veyrat2023,PhysRevB.110.054504,veyrat2024dissipationless}. In the following, we show that: (i) these two sets map to distinct nodal structures in the zero-SOC limit—one emerging from mirror-protected nodal lines and the other from orbital Weyl nodes; and
(ii) both sets share a common underlying origin: band folding effects.

 \begin{figure*}[t]
    \centering
    \includegraphics[width=0.55\columnwidth]{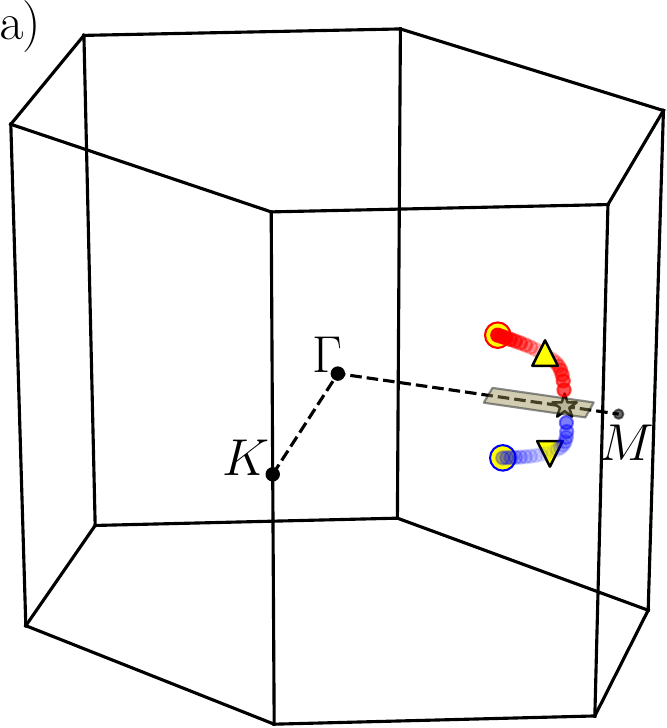}
    \includegraphics[width=0.7\columnwidth]{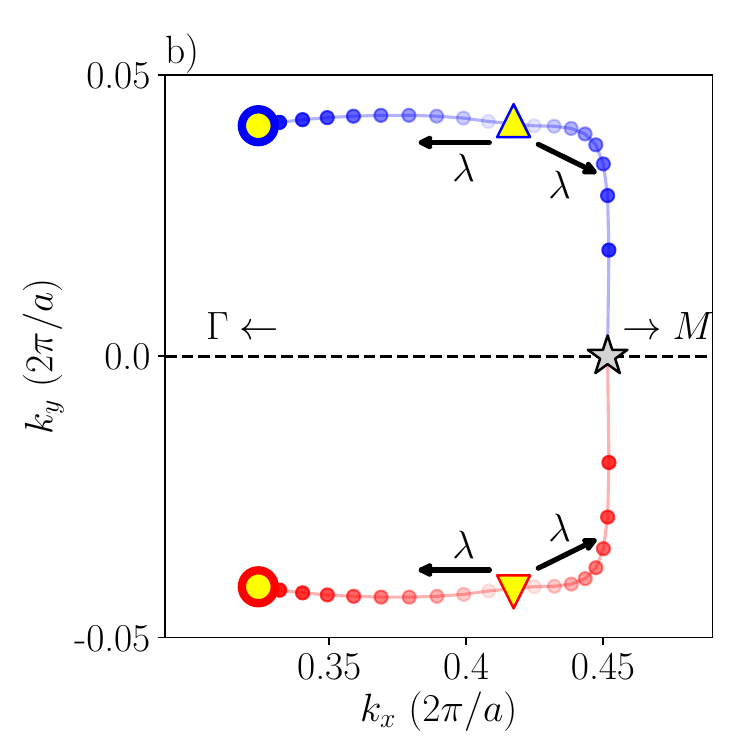}
    \includegraphics[width=0.7\columnwidth]{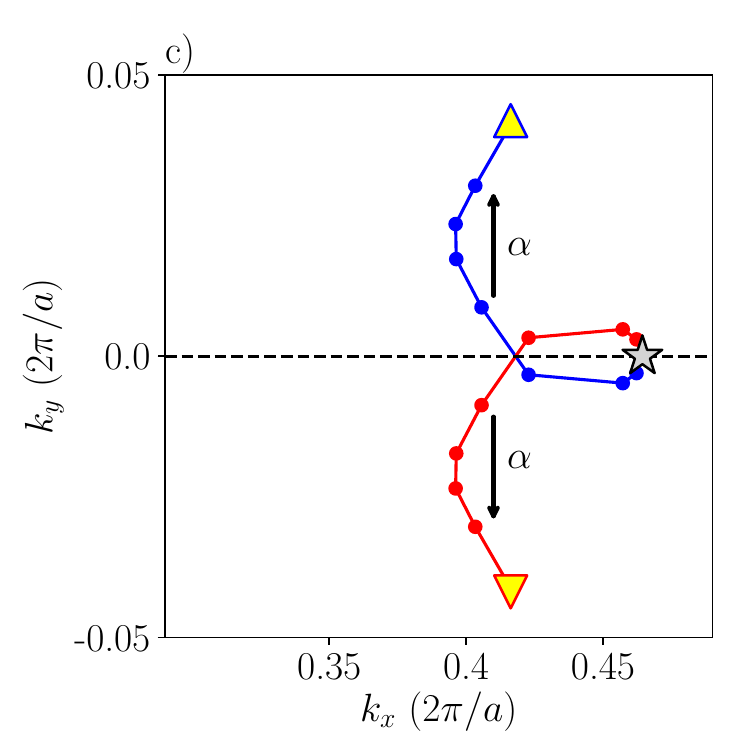}
    \caption{ (a) Evolution of the Weyl nodes closest to the Fermi energy as a function of the spin-orbit coupling ($\lambda$ in Eq. \ref{eq_interpolation2}). Red and blue correspond to different Chern number signs. For $\lambda=0$, two Weyl nodes of identical Chern number overlay at the yellow triangles. Each of these corresponds to a monopole of the orbital Berry curvature. The yellow circles correspond to the fully relativistic limit.  The same phenomenology occurs at $C_3$-related planes and, due to time-reversal symmetry, at opposite $\mathbf{k}$.
    (b) Projection of the nodal points onto the $k_x$-$k_y$ plane. The area covered by the graphic corresponds to the small rectangle shown in panel (a). Dots with lighter shades correspond to weaker spin-orbit coupling.
    (c) For $\lambda=0$, trajectory of the Weyl nodes as a function of the crystalline distortion ($\alpha$ in Eq. \ref{eq_interpolation}).  At $\alpha\approx0.39$, a pair of Weyl nodes emerges from the $\Gamma$-$M$ line (gray star). The arrows indicate the sense of increasing $\lambda$ (b) or $\alpha$ (c).}
    \label{fig_wp_traj}
\end{figure*}

\begin{figure}[h]
    \centering
    \includegraphics[width=\columnwidth]{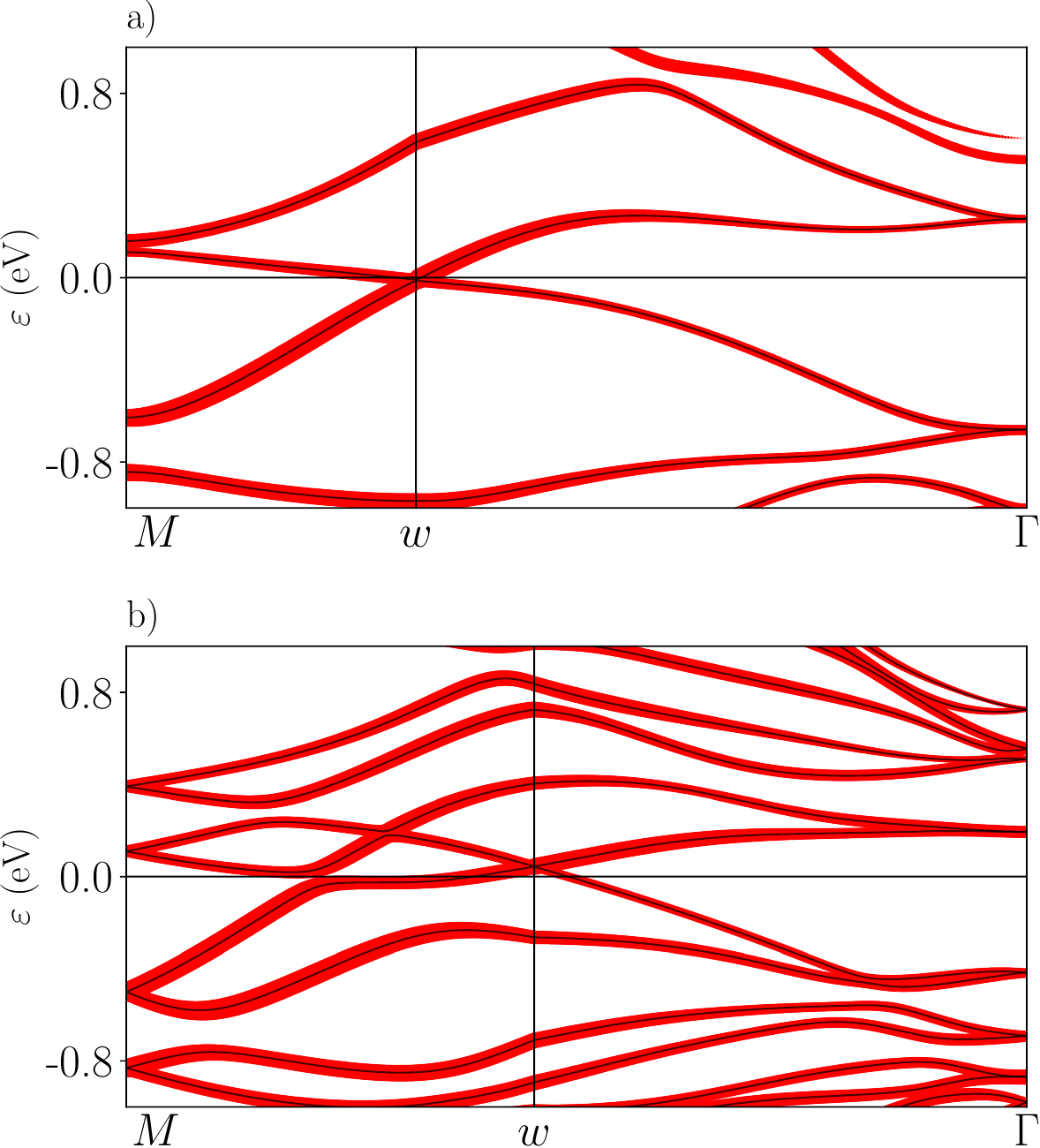}
    \caption{Band structure of PtBi$_2$ along one of the Weyl nodes without SOC (a) or with SOC (b). In the first case, the point $w=(0.416,-0.041,0.124)2\pi/a$ while in the second $w=(0.324,-0.041,0.153)2\pi/a$. The width of the red lines measures the unfolding weight, which tends to zero for folded bands.}
    \label{fig_wn_folding}
\end{figure}

\subsection{Set I of Weyl nodes}

An important characteristic of these nodes is that they are already present in the electronic structure in the absence of SOC.
 To demonstrate this point, we use the interpolation defined by Eq. (\ref{eq_interpolation2}) and follow the evolution of the nodes with $\lambda$.  
Figure \ref{fig_wp_traj}a) illustrates the trajectories of a pair of Weyl nodes in the three-dimensional BZ, while Fig.~\ref{fig_wp_traj}b) presents the projection onto the $k_x$--$k_y$ plane. Yellow disks are used to indicate the positions of the nodes in the fully relativistic limit $\lambda = 1$.
As $\lambda$ decreases, these nodes shift in energy and momentum while a pair of opposite-chirality nodes emerges from the $\Gamma$-$M$ (gray star). Upon further reduction of $\lambda$, nodes of the same chirality approach each other and, in the limit of vanishing SOC,  overlay in positions depicted as yellow triangles.

By integrating the orbital Berry curvature over a surface that encloses one node at $\lambda = 0$, we find that the crossing of each spin sector carries a Chern number of one. This situation resembles the calculations of WTe$_2$~\cite{soluyanov2015type} and of TaIrTe$_4$~\cite{PhysRevB.93.201101} and contrasts with semimetals where the Weyl nodes stem from nodal lines upon including SOC. In PtBi$_2$, the $I$-breaking crystalline distortion induces the emergence of these nodes, which are created pairwise at the  $\Gamma$-$M$ line at  $\alpha\sim0.39$ (Fig. \ref{fig_wp_traj}c).

Figure~\ref{fig_wn_folding} presents the band structure across one of the nodes, obtained either in the absence of SOC (a) or including SOC (b). While SOC naturally introduces a Rashba-like splitting of the bands, it is evident that, with or without SOC,  one of the bands forming the Weyl node connects smoothly to the occupied states at $\Gamma$ closest to the Fermi energy. As discussed previously (see Fig.~\ref{fig_dft}), these states belong to the occupied manifold originally folded from $K_{nc}$ to $\Gamma$ and subsequently split by the crystalline distortion. This shows that the band folding leaves a clear fingerprint in the structure of the Weyl nodes.

\begin{figure}[t]
    \centering
    \includegraphics[width=0.65\columnwidth]{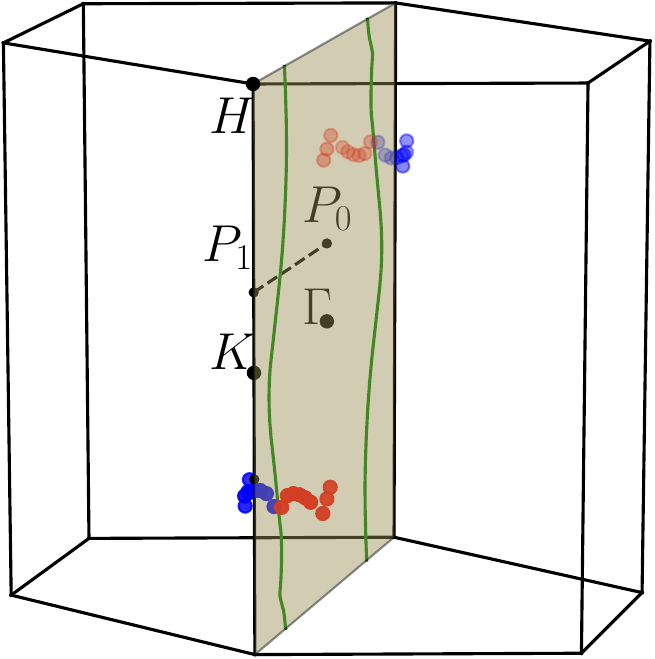}
    \caption{Evolution of the Set II of Weyl nodes
    as a function of the spin-orbit coupling [$\lambda$ in Eq. (\ref{eq_interpolation2})]. Red and blue correspond to different Chern number signs. The green lines are contained in the highlighted reflection-invariant plane and correspond to nodal lines in the absence of spin-orbit coupling. The same phenomenology occurs at $C_3$-related planes. }
    \label{fig_wp_traj_mirror}
\end{figure}

The fact that spin degrees of freedom are not essential for the existence of the Weyl nodes is consistent with the smooth spin texture around the nodes found in Ref.~\cite{PhysRevB.110.054504}. As we have here shown, orbital physics associated with the crystalline distortion is at the origin of these Weyl nodes, with the SOC being relevant for the precise number of nodes, their final location in energy and momentum space, and, as it is shown in Sec. \ref{sec_fa}, the aspect and connectivity of the associated topological Fermi arcs.

\subsection{Set II of Weyl nodes}

Fig. \ref{fig_wp_traj_mirror} displays, using green curves, nodal lines that lie within the mirror-invariant planes $\Gamma$-$K$-$H$. 
Upon inclusion of SOC, these nodal lines gap out leaving behind Weyl nodes near them. This is the mechanism that gives rise to the Set II of Weyl nodes in PtBi$_2$. The blue and red dots in Fig. \ref{fig_wp_traj_mirror} illustrate the trajectory of opposite-chirality Weyl nodes as a function of $\lambda$.

\begin{figure}[t]
    \centering
    \includegraphics[width=\columnwidth]{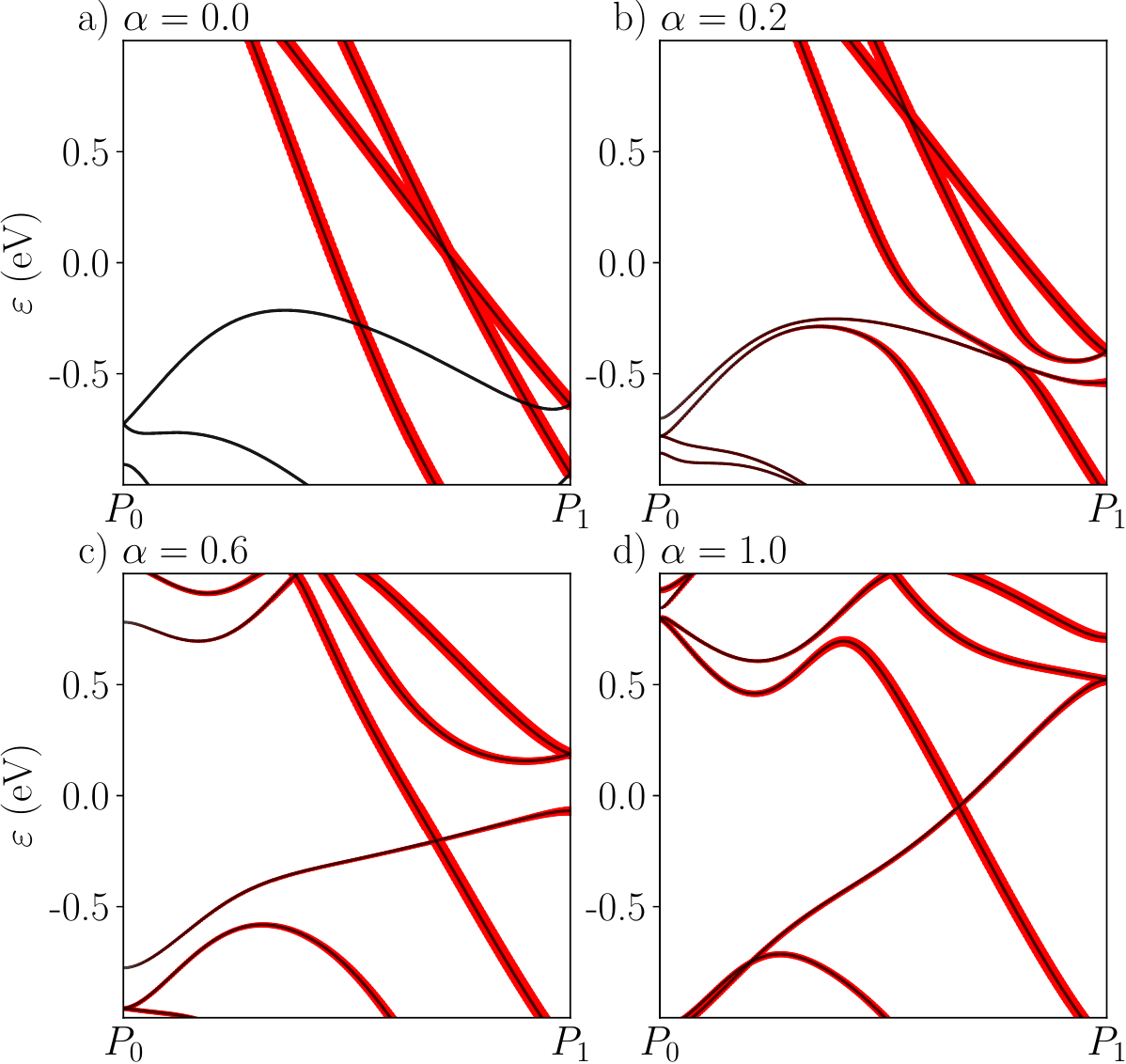}
	\caption{a-d) Band structure of PtBi$_2$ in the absence of spin-orbit coupling for different crystal structures. The parameter $\alpha$ interpolates between the centrosymmetric limit ($\alpha=0$) and the noncentrosymmetric case ($\alpha=1$). The $\mathbf{k}$-path is indicated in Fig. \ref{fig_wp_traj_mirror}, it corresponds to $k_z=0.15\times2\pi/a$. The size of the red dots measures the unfolding weight, which tends to zero for folded bands, so that in panel a) these appear as black curves.}
    \label{fig_nl_w}
\end{figure}

Interestingly, a calculation of the nodal lines as a function of the crystalline distortion indicates that their origin is also related to the band folding. This can be seen more directly by inspecting the bands along the paths where the position of the nodal line depends weakly on $\alpha$. The band structure along one such path is shown in Fig. \ref{fig_nl_w}a-d) for various $\alpha$. 

At the $I$-symmetric limit $\alpha = 0$, various crossings between folded and nonfolded bands can be observed. Importantly, in this limit, the crossings involve three bands (not counting spin). This occurs because the reflection symmetry causes two distinct states to be folded onto the $\Gamma$-$K$ path, as can be observed in the BZs shown in Fig.~\ref{fig_dft}. The two folded states, degenerate in this limit, have opposite mirror eigenvalue.
Such three-fold crossings are unstable to perturbations that reduce translational symmetry; however, the crystalline distortion in PtBi$_2$ preserves the reflection symmetry. Since the two folded bands correspond to states of opposite mirror eigenvalues, necessarily one of them cannot hybridize with the nonfolded band involved in the three-fold crossing. Consequently, for finite $\alpha$, a two-fold crossing persists in the distorted structure. 

Let us summarize the two main conclusions of this section. First, 
the underlying topology of the nodal structure in the absence of SOC differs between the Weyl nodes of different energy: one set follows the nodal-line mechanism while the set closer to the Fermi energy originates from monopoles of the orbital Berry curvature.
Second, the reduction of translational symmetry due to the inversion-breaking crystalline distortion plays a crucial role in the formation of the two sets of Weyl nodes in PtBi$_2$.  

\section{Spin-orbit coupling and Fermi arcs}
\label{sec_fa}

We have shown that the Weyl nodes closest to the Fermi energy (set I in Section \ref{sec_nl}) also appear in the absence of SOC, originating from orbital physics associated with the reduced translational symmetry. In the bulk electronic structure, the primary role of SOC is to couple and energetically split Weyl nodes of opposite spin, leading to pairwise annihilation processes and determining the final positions of the surviving nodes. In this context, it is of particular interest to examine how the surface Fermi arcs evolve as SOC is varied.

To this aim, we calculate the surface electronic structure of a semi-infinite slab based on the Wannier tight-binding model, using the methodology of Ref.~\cite{sancho1985highly} as implemented in Ref.~\cite{koepernik23}.

\begin{figure}[t]
    \centering
    \includegraphics[width=\columnwidth]{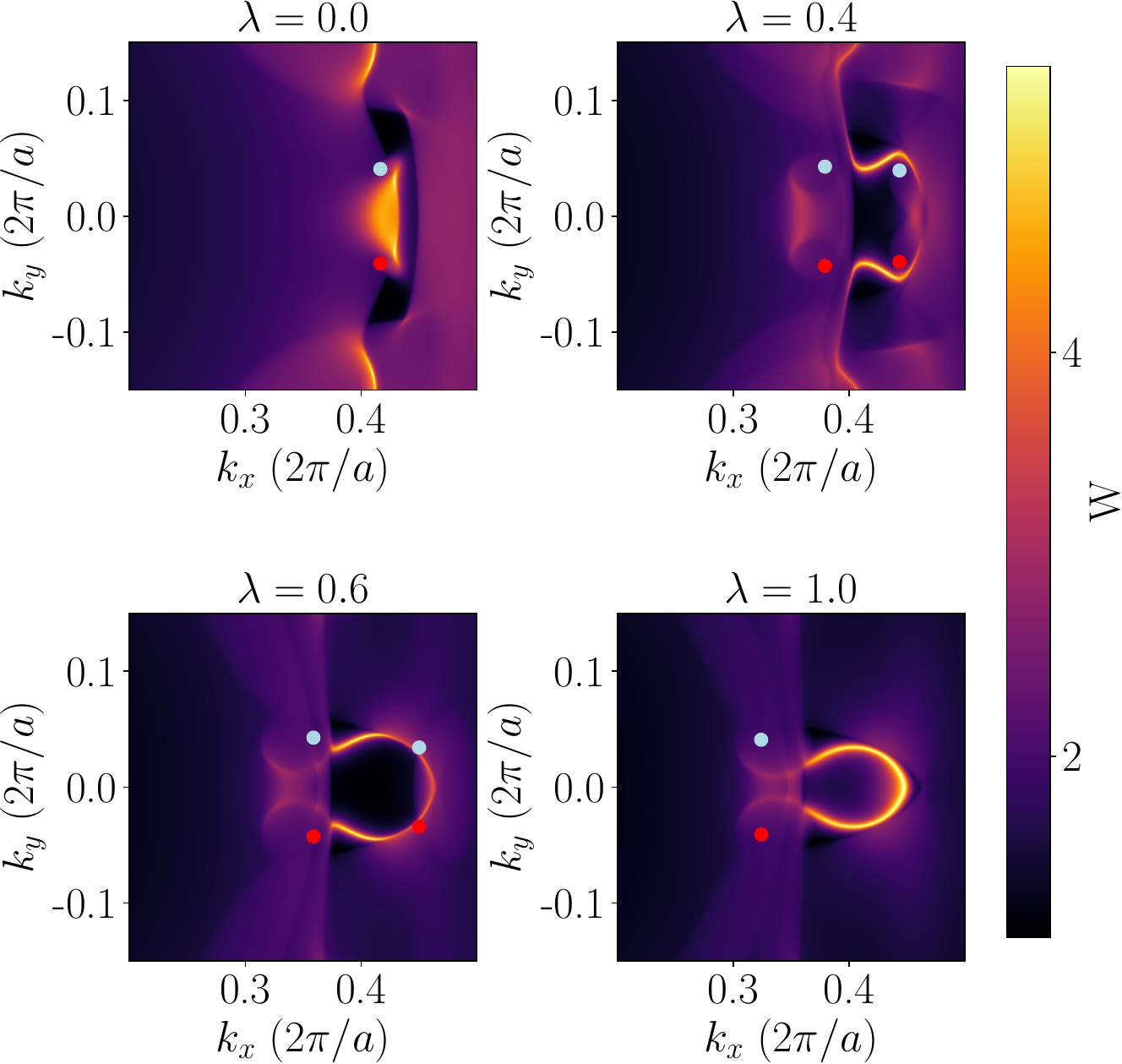}
    \caption{Surface spectral weight $A(\mathbf{k})$ at the Fermi energy for different values of the spin-orbit coupling strength, parameterized by $\lambda$ in Eq.~\ref{eq_interpolation2}. The colormap is based on the weight $W(\mathbf{k})=[A(\mathbf{k})]^{1/3}$, which enhances visualization of the (weaker) spectral contributions arising from projected bulk states.}
    \label{fig_fermiarcs}
\end{figure}

Figure~\ref{fig_fermiarcs} shows the surface spectral density at the Fermi energy for a surface terminated in the decorated honeycomb layer. The color scale represents the spectral weight integrated over a depth of $10$\,\AA\ beneath the surface. 
In the absence of SOC ($\lambda = 0$), a bright Fermi arc is clearly visible, connecting the projections of Weyl nodes with opposite chirality across the $\Gamma$–$M$ line. Notice that because of SU(2) symmetry, these Fermi arcs have spin degeneracy.

As $\lambda$ increases, Weyl nodes of opposite spin couple and shift in momentum space. Accordingly, the Fermi arcs evolve, along with the spectral weight carried by bulk states, which can be seen to mediate the spectral flow between distinct pairs of nodes. In the fully relativistic limit ($\lambda = 1$), a well-defined Fermi arc is once again observed.
While the SOC is not instrumental for the existence of the topological surface Fermi arcs, their precise shape and extent differ noticeably from those in the $\lambda = 0$ case.

\vspace{5mm}
\section{Conclusions}
\label{sec_conclusions} 

We have discussed the electronic structure of trigonal PtBi$_2$, aiming to correlate its salient characteristics with its crystal structure. Our main results follow from comparisons with a centrosymmetric reference limit, which has been considered in earlier density-functional calculations~\cite{PhysRevB.110.054504} and is experimentally achievable via moderate substitution of Bi by Te or Sb~\cite{Takaki2022}.

One important finding of this work is the identification of a strong electronic reconstruction associated with the inversion-breaking crystalline distortion. In particular, we have demonstrated that this reconstruction is fundamentally tied to the fact that the distortion also breaks translational symmetries. 
The strong reduction of the spectral weight at the Fermi energy can naturally be experimentally tested. Encouragingly, we note that the observed superconducting critical temperature is higher in centrosymmetric samples~\cite{Takaki2022}. Making a clear statement, however, requires a careful study, as surface superconductivity effects could complicate direct comparisons~\cite{Veyrat2023,Kuibarov2023,schimmel2024surface,PhysRevB.108.165144,trama2024self}.

The low-energy electronic reconstruction can be understood in analogy with the Peierls transition in one dimension, with notable key differences arising due to the higher dimensionality of the present case and of the interplay between broken inversion and reduced translational symmetries~\cite{us_short}. 
While the crystalline distortion strongly suppresses the spectral weight at the Fermi energy, codimensionality arguments for an inversion-broken three-dimensional system dictate that the distorted phase can stabilize a Weyl semimetallic phase~\cite{Murakami_2007}. 
We have shown that different sets of Weyl nodes in PtBi$_2$ reflect different mechanisms in which the opening of a full gap can become obstructed.

An important direction for future study is the role of phonons. A characterization of their interplay with the electronic structure could provide key insights into the stabilization mechanism of the noncentrosymmetric distortion. In particular, the identification of a strong electronic reconstruction associated with the crystalline distortion motivates a detailed study of the phonon spectrum and of the electron-phonon coupling in both centrosymmetric and noncentrosymmetric phases. 

Lastly, our findings can be of interest in contexts where translational symmetry is spontaneously reduced and spectral weight suppressed at the Fermi energy, such as in antiferromagnetic transitions. When the resulting distorted phase lacks inversion symmetry, the resulting electronic structure can naturally be semimetallic rather than insulating. 

\section{Appendix: Phonon spectrum of the centrosymmetric limit}
\label{sec_app}

In this Appendix, we present first-principles calculations of the phonon spectrum for the centrosymmetric (CS) trigonal structure PtBi$_2$. These results provide insight into the metastability of this phase and the possible existence of a transition pathway toward the noncentrosymmetric (NCS) structure.

DFT calculations were performed within the GGA, as implemented in the \textsc{Quantum ESPRESSO} package~\cite{QE-2009,QE-2017}. The calculations employed a plane-wave basis set with kinetic-energy cutoffs of 60\,Ry for the wavefunctions and 600\,Ry for the charge density. Ultrasoft pseudopotentials were used for both Pt and Bi atoms. The Brillouin zone was sampled using a $16 \times 16 \times 12$ Monkhorst–Pack $\mathbf{k}$-point grid. We adopted the Marzari–Vanderbilt smearing scheme~\cite{PhysRevLett.82.3296} with a broadening parameter of $\delta = 0.27$\,eV. A full optimization of both the atomic positions and lattice parameters was carried out using the BFGS algorithm, with convergence thresholds of $10^{-5}$\,Ry for the total energy and $2 \times 10^{-6}$\,Ry/Bohr for the forces. Phonon spectra were computed within density-functional perturbation theory on an $8 \times 8 \times 6$ $\mathbf{q}$-point grid. We verified at selected $\mathbf{q}$ points that the results depend only weakly on $\delta$. Dynamical matrices were computed using a self-consistency threshold of $10^{-14}$.

\begin{figure}[h]
    \centering
    \includegraphics[width=\columnwidth]{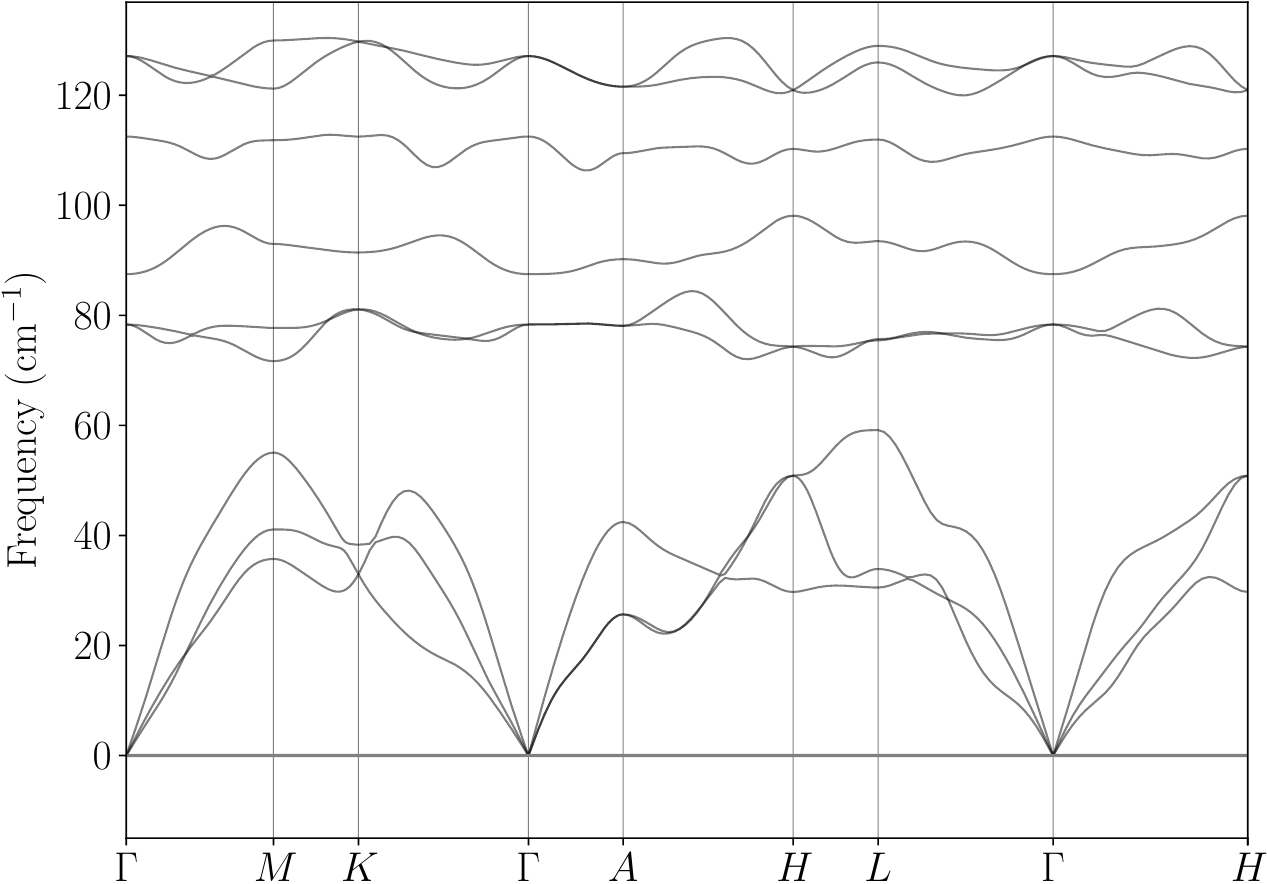}
    \caption{Phonon spectrum calculated for PtBi$_2$ in space group P$\bar{3}$m1. No unstable (imaginary) modes are observed.}
    \label{fig_phonons}
\end{figure}

Figure~\ref{fig_phonons} shows the resulting phonon dispersion. 
We have checked by calculation of the density of states that oscillation frequencies are real in the full Brillouin zone.
The absence of imaginary frequencies indicates that the CS structure is dynamically stable at zero temperature. This suggests that the phase is metastable, with a local minimum in the energy landscape.

This structure, however, has not been observed experimentally. This suggests that the energy barrier stabilizing the CS phase may be too small, or that the synthesis conditions do not favor trapping the system in the CS configuration. In particular, the dynamics of the synthesis process, such as cooling rates, chemical potential gradients, or kinetic pathways, may prevent access to this metastable state, even if it is locally stable.
A centrosymmetric structure has been reported in Te-substituted samples~\cite{Takaki2022}. Although the mechanism by which Te stabilizes this phase remains to be fully understood, such doped systems may offer a useful platform to explore transitions between the CS and NCS phases experimentally.

The existence of a metastable CS structure implies that external parameters, such as pressure, strain, or chemical substitution, could in principle be used to tune the system between phases. 
FPLO calculations for fully relaxed crystal structures yield an energy difference $E_{CS}-E_{NCS}\sim 52\,$meV.
However, since PtBi$_2$ exhibits multiple known polymorphisms (including cubic and orthorhombic structures), the structural evolution upon destabilization of the CS phase may not necessarily lead directly to the trigonal NCS phase.

\section{Acknowledgements}
JF acknowledges useful discussions with Riccardo Vocaturo, Klaus Koepernik, Oleg Janson, Ion Cosma Fulga, and Jeroen van den Brink.
Computational resources were provided by the HPC cluster of the Physics Department at Centro Atómico Bariloche (CNEA).

\bibliographystyle{apsrev4-2}
\bibliography{PtBi2.bib}
\end{document}